# Magnetic Transitions under Ultrahigh Magnetic Fields of up to 130 T in the Breathing Pyrochlore Antiferromagnet LiInCr$_4$O$_8$

Yoshihiko Okamoto[1,*], Daisuke Nakamura[2], Atsushi Miyake[2], Shojiro Takeyama[2], Masashi Tokunaga[2],
Akira Matsuo[2], Koichi Kindo[2], and Zenji Hiroi[2]

[1]*Department of Applied Physics, Nagoya University, Nagoya 464-8603, Japan*
[2] *Institute for Solid State Physics, University of Tokyo, Kashiwa 277-8581, Japan*

The magnetization processes of the spin-3/2 antiferromagnet LiInCr$_4$O$_8$ comprising a "breathing" pyrochlore lattice, which is an alternating array of small and large tetrahedra, are studied under ultrahigh magnetic fields of up to 130 T using state-of-the-art pulsed magnets. A half magnetization plateau is observed above 90 T to 130 T, suggesting that LiInCr$_4$O$_8$ has a strong spin–lattice coupling, similar to conventional chromium spinel oxides. The magnetization of LiGa$_{0.125}$In$_{0.875}$Cr$_4$O$_8$, in which the structural and magnetic transitions at low temperatures have been completely suppressed, shows a sudden increase above 13 T, indicating that a spin gap of 2.2 meV exists between a tetramer singlet ground state and an excited state with total spin 1, with the latter being stabilized by the application of a magnetic field. The breathing pyrochlore antiferromagnet is found to be a unique frustrated system with strong spin–lattice coupling and bond alternation.

## I. INTRODUCTION

The chromium spinel oxide ACr$_2$O$_4$ with a nonmagnetic A$^{2+}$ ion, such as Zn$^{2+}$, Mg$^{2+}$, Cd$^{2+}$, or Hg$^{2+}$, is one of the most intensively studied geometrically frustrated magnets [1]. Cr$^{3+}$ ions with three localized 3$d$ electrons carrying an $S$ = 3/2 Heisenberg spin form a pyrochlore lattice, in which antiferromagnetic interactions are dominant. ACr$_2$O$_4$ undergoes a long-range magnetic order with a complex spin structure, accompanied by a structural distortion, although the transition temperature is lowered due to geometrical frustration of the pyrochlore lattice [2-4]. Another distinctive feature of ACr$_2$O$_4$ is its magnetic phase transitions induced by applying a magnetic field. ACr$_2$O$_4$ with A = Cd and Hg exhibits magnetic transitions at magnetic fields of 28 and 10 T, respectively [5,6]. Above this field, magnetization curves show a plateau at half the saturation magnetization, $M_s$, known as the half magnetization plateau. Neutron diffraction measurements under magnetic fields revealed a common magnetic structure at the plateau phase, namely, a ferrimagnetic order maintaining cubic crystal symmetry with each tetrahedron having a 3-up-1-down collinear spin configuration [7,8]. ACr$_2$O$_4$ single crystals with A = Zn and Mg are also found to show half magnetization plateaus above approximately 140 T [9-12]. Common to the aforementioned four spinel oxides, half magnetization plateaus are formed in wide magnetic-field ranges of several tens of tesla [6,13-16], in contrast to the narrow ones observed in Ising pyrochlore magnets, such as Ho$_2$Ti$_2$O$_7$ [17]. Theoretical studies suggest the important role of spin–lattice coupling in stabilizing the wide magnetization plateaus in ACr$_2$O$_4$ [18,19].

Here we focus on an A-site ordered Cr spinel oxide LiInCr$_4$O$_8$. In LiInCr$_4$O$_8$, Li$^+$ and In$^{3+}$ ions form a zinc-blende type order [20,21]. This atomic order causes chemical pressure on the Cr$^{3+}$ pyrochlore lattice, resulting in an alternation in the size of adjacent Cr$_4$ tetrahedra, as shown in the inset of Fig. 1(a), and a reduction in crystal symmetry from cubic $Fd$–3$m$ to another cubic space group, $F$–43$m$. This bond-alternated pyrochlore lattice is known as the "breathing" pyrochlore lattice [21]. There are antiferromagnetic couplings between neighboring Cr$^{3+}$ spins, as indicated by a negative Weiss temperature of $\theta_W = -332$ K [21]. The magnitudes of antiferromagnetic interactions on the small and large tetrahedra, $J$ and $J'$, respectively, are quite different. We estimated the ratio of $J$ and $J'$, defined as the breathing factor, $B_f = J'/J$, to be approximately 0.1 for LiInCr$_4$O$_8$ deduced from an empirical relationship between the strength of magnetic interactions and the Cr–Cr distances [21]. This small $B_f$ indicates that LiInCr$_4$O$_8$ lies close to the isolated tetrahedra limit ($B_f = 0$) rather than the uniform pyrochlore limit ($B_f = 1$).

LiInCr$_4$O$_8$ shows spin-gap behavior below 65 K with the magnetic susceptibility strongly decreasing with decreasing temperature. In this temperature region, four spins on a small tetrahedron form a tetramer singlet with the total spin on a small tetrahedron $S_t = 0$. As discussed in the studies of the pseudospin-1/2 breathing pyrochlore antiferromagnet Ba$_3$Yb$_2$Zn$_5$O$_{11}$, the tetramer singlet has a ground state degeneracy caused by the tetrahedral symmetry [22-24]. At approximately 15 K, LiInCr$_4$O$_8$ is found to exhibit successive structural and magnetic phase transitions and go into an antiferromagnetically ordered ground state on the distorted breathing pyrochlore lattice, suggestive of the presence of a strong spin–lattice coupling similar to the case of ACr$_2$O$_4$ [25-27].

These phase transitions in LiInCr$_4$O$_8$ are suppressed by Ga substitutions. LiGa$_{0.125}$In$_{0.875}$Cr$_4$O$_8$ shows a similar magnetic susceptibility to that of LiInCr$_4$O$_8$, suggesting that the magnetic interactions are almost identical [28]. However, there is no peak in the heat capacity data of LiGa$_{0.125}$In$_{0.875}$Cr$_4$O$_8$ down to 0.5 K, indicating that no phase transitions take place. Moreover, the magnetic susceptibility shows spin-gap behavior continuing down to 2 K without glass-type hysteresis. This is probably because the structural transition has been suppressed by the substitution, which enables us to study the intrinsic properties of the breathing pyrochlore antiferromagnets, such as the effects of high magnetic field on the tetramer singlet, different from LiInCr$_4$O$_8$.

In this study, we report state-of-the-arts high-field magnetization measurements on powder samples of LiInCr$_4$O$_8$ and LiGa$_{0.125}$In$_{0.875}$Cr$_4$O$_8$ using nondestructive and destructive pulsed magnets. They show significantly different magnetization processes reflecting the different zero-field magnetic states. The magnetization of LiInCr$_4$O$_8$ monotonically increases with increasing magnetic fields up to 72 T, which is the highest magnetic field measured by the nondestructive pulsed magnet. In contrast, that of LiGa$_{0.125}$In$_{0.875}$Cr$_4$O$_8$ is smaller in a low magnetic field region and strongly increases above 13 T, corresponding to the closing of the spin gap. Moreover, we discovered a half magnetization plateau in the magnetization curve of LiInCr$_4$O$_8$ above 90 T. This finding indicates the presence of strong spin–lattice coupling in LiInCr$_4$O$_8$.

## II. EXPERIMENTAL DETAILS

High-resolution magnetization measurements up to 72 T on powder samples of LiInCr$_4$O$_8$ and LiGa$_{0.125}$In$_{0.875}$Cr$_4$O$_8$ were performed using a multilayered nondestructive pulsed magnet with a duration time of 4 ms. Each sample was prepared from the same batch as used in Ref. 28. These samples show the sharp diffraction peaks in powder X-ray and neutron diffraction patterns, indicative of the good crystallinity [21,28]. The very small Curie tails in magnetic susceptibility data suggest that the surface states and lattice defects, giving rise to the orphan spins, in the samples are quite few [28]. The magnetizations were measured at 1.4 K by the electro-magnetic induction method employing a coaxial pick-up coil. Since it is difficult to obtain the absolute values of magnetization using this method, we have calibrated the data to fit other magnetization curves measured on the same samples up to 7 T using a Magnetic Property Measurement System (Quantum Design).

A magnetization process of LiInCr$_4$O$_8$ at ultrahigh magnetic fields of up to 130 T was measured using a destructive single-turn-coil megagauss generator equipped with 200 kJ and 50 kV fast-condenser banks [29]. The sample used in this measurement is identical to that used in the above measurements. The pulse field duration is about 6.5 μs and the magnetic field reaches its maximum value in approximately 2.5 μs. The magnetization curve was obtained by the electro-magnetic induction method with a co-axial self-compensated magnetic pick-up coil. Details of the measurement techniques will be reported elsewhere, but their essence is similar to those reported in Ref. 30, except the configuration of the pick-up coil (changed from a parallel pair to co-axial). The pick-up coil was set in a He-flow type cryostat solely made of glass-epoxy (FRP, G-10) with special low-temperature glue (Nitofix SK-229, NITTO DENKO Co. Ltd.). The cryostat with its outermost tube diameter of 7.0 mm is inserted precisely into the 12 mm bore single-turn coil. Pairs of measurements with and without the sample in the identical coil under the exactly same discharging conditions were carried out to eliminate background noises superposed on the intrinsic signal from the sample.

## III. RESULTS AND DISCUSSION

### A. Magnetization processes measured by employing a nondestructive pulsed magnet

Figure 1(a) shows magnetization curves of a LiInCr$_4$O$_8$ powder sample measured up to 72 T at 1.4 K. The data measured up to 15, 58, and 72 T completely overlap with each other. The magnetization $M$ increases with increasing magnetic field $H$ and reaches 0.49 μ$_B$/Cr at 72 T. This $M$ corresponds to $0.17M_s$, as shown in Fig. 2, providing a saturation magnetization of $M_s = gS\mu_B = 2.964$ μ$_B$/Cr, estimated from the Lande g factor, $g = 1.976$, determined by electron spin resonance experiments [27]. As seen in the top panel of Fig. 2, the $dM/dH$ of LiInCr$_4$O$_8$ shows a small peak at 7 T, indicating that the slope of the $M$–$H$ curve takes a local maximum at this $H$, where there may be a small change in the magnetic structure. Above 20 T, the $M$–$H$ curves of LiInCr$_4$O$_8$ show slightly concave-upward behavior, probably related to the increase of $M$ towards a half magnetization plateau, as discussed later.

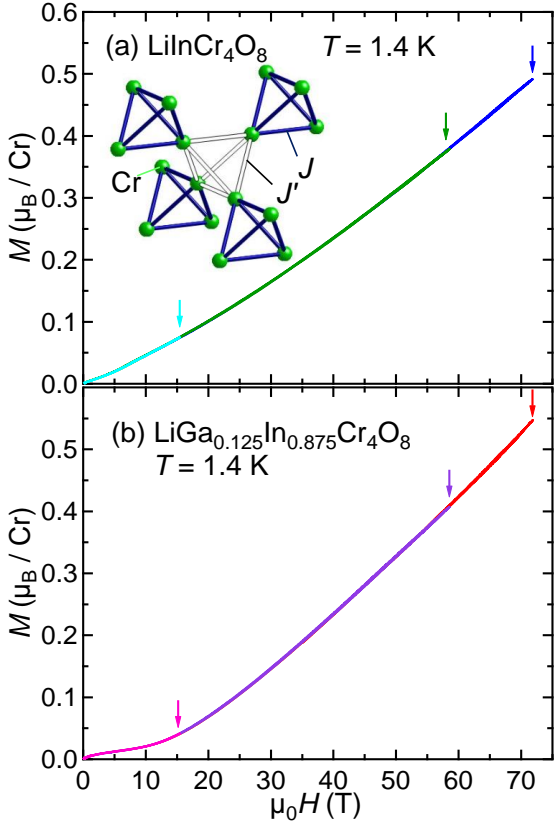

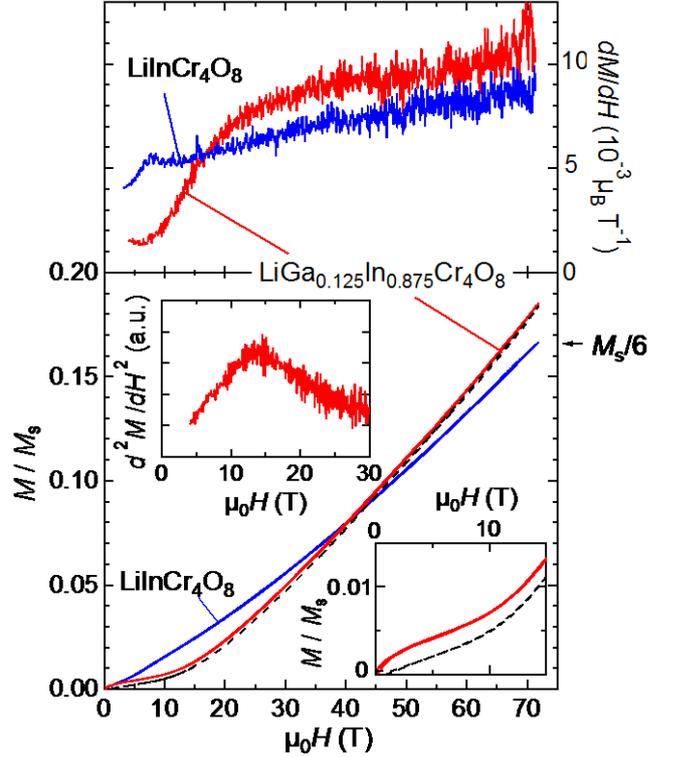

Fig. 1. Magnetization curves of powder samples of (a) $LiInCr_4O_8$ and (b) $LiGa_{0.125}In_{0.875}Cr_4O_8$. Measurements were performed up to 15, 58, and 72 T at 1.4 K using a multilayered pulsed magnet. The highest magnetic field for each measurement is indicated by an arrow. The inset shows a breathing pyrochlore lattice made of $Cr^{3+}$ ions.

Fig. 2. Normalized magnetization curves of powder samples of $LiInCr_4O_8$ and $LiGa_{0.125}In_{0.875}Cr_4O_8$ measured up to 72 T at 1.4 K. The data for $LiGa_{0.125}In_{0.875}Cr_4O_8$ after the removal of the orphan spin contribution are shown as a dotted curve. $dM/dH$ of the $M$–$H$ curves and $d^2M/dH^2$ of that of $LiGa_{0.125}In_{0.875}Cr_4O_8$ are shown at the top and in the upper inset, respectively. The lower inset shows the low magnetic-field region of the $LiGa_{0.125}In_{0.875}Cr_4O_8$ data.

The magnetization process of $LiGa_{0.125}In_{0.875}Cr_4O_8$ is significantly different from that of $LiInCr_4O_8$. As shown in Fig. 1(b), the magnetization curves of a $LiGa_{0.125}In_{0.875}Cr_4O_8$ powder sample measured up to 15, 58, and 72 T completely overlap with each other. The $M$–$H$ curve of $LiGa_{0.125}In_{0.875}Cr_4O_8$ is concave downward below 10 T. This behavior suggests the presence of nearly free spins, which are most likely orphan spins appearing around crystal defects of the spin-singlet tetramers [31]. The number of orphan spins are estimated by fitting the magnetization curve to the Brillouin function for $S = 3/2$; 0.26% of all $Cr^{3+}$ spins. This value is nearly equal to 0.2%, as estimated by the Curie–Weiss fit of the temperature dependence of magnetic susceptibility [28]. The magnetization of $LiGa_{0.125}In_{0.875}Cr_4O_8$ after subtracting the orphan spin contribution is very small in the low magnetic field region, as shown in Fig. 2, suggesting that $LiGa_{0.125}In_{0.875}Cr_4O_8$ remains in the tetramer singlet state at 1.4 K.

The magnetization of $LiGa_{0.125}In_{0.875}Cr_4O_8$ rapidly increases with increasing magnetic field above approximately 13 T. As shown in the inset of Fig. 2, the $d^2M/dH^2$ of $LiGa_{0.125}In_{0.875}Cr_4O_8$ shows a maximum at 13 T, meaning that the $M$ increases most steeply at this $H$. This steep increase reflects the fact that the energy gap $\Delta$ between the tetramer singlet state with $S_t = 0$ and the excited state with $S_t = 1$ becomes zero when applying the magnetic field. The energy scale of the $\mu_0H = 13$ T corresponds to $\Delta = 2.2$ meV ($\Delta/k_B = 26$ K), given the values of $g$ and $S$ in this compound. This energy scale is not far from the energy gap of $\Delta/k_B \sim 30$ K, estimated from the temperature dependence of $1/T_1$ of $^7$Li-NMR in the paramagnetic phase of $LiInCr_4O_8$ [25], suggestive of the same origin.

Above 13 T, the $M$ of $LiGa_{0.125}In_{0.875}Cr_4O_8$ monotonically increases with increasing $H$ and reaches 0.55 $\mu_B$/Cr = $0.19M_s$ at 72 T. This $M$ is already larger than $M_s/6$, which is the magnetization when $S_t = 1$ spins are fully polarized along the applied magnetic field. In general, breathing pyrochlore antiferromagnets with small $B_f$ are expected to show stepwise *M–H* curves at sufficiently low temperature, in which plateaus appear at $M = nM_s/4S$, where $n$ is an integer between 0 and $4S$, the same as in the isolated tetrahedra case. In fact, the *M–H* curve of the pseudospin-1/2 $Ba_3Yb_2Zn_5O_{11}$ with $J/k_B = 7$ K measured at 0.5 K is stepwise and shows a plateau at $M_s/2$ [23], which remain as anomalies corresponding to the closing of the spin gap and the transition to the plateau in 1.8 K data [22]. It is not clear why there is no signature corresponding to the $M_s/6$ plateau in the $LiGa_{0.125}In_{0.875}Cr_4O_8$ data measured at 1.4 K, which has a much smaller energy scale than $J/k_B = 60$ K [28]. The $J'$ larger than that in $Ba_3Yb_2Zn_5O_{11}$ may have an important role in the absence of the $M_s/6$ plateau in $LiGa_{0.125}In_{0.875}Cr_4O_8$.

## B. Magnetization processes measured by the single-turn coil method

Figure 3 shows a magnetization process of a $LiInCr_4O_8$ powder sample measured up to 130 T at approximately 10 K by the single-turn coil method. The $M$ jumps to 1.5 $\mu_B$/Cr at approximately 90 T, indicating a magnetic transition occurs at this $H$. From the transition field to the highest measured field of 130 T, the $M$ is almost constant at 1.5 $\mu_B$/Cr. This $M$ corresponds to $0.50M_s$, given the values of $g$ and $S$ of $LiInCr_4O_8$, indicating that $LiInCr_4O_8$ is in a half magnetization plateau state in this magnetic field region. A closer look at the magnetization process shown in Fig. 3 reveals that the jumps of $M$ occur at 95–100 T and 85–90 T with increasing and decreasing $H$, respectively, indicative of the presence of hysteresis in this transition. This result suggests that the magnetic transition to the half magnetization plateau occurs as a first-order phase transition.

The half-magnetization plateau formed over a wide $H$ region and the first-order transition to the plateau are identical to the behavior of $ACr_2O_4$. These features strongly suggest for the $M_s/2$ plateau phase of $LiInCr_4O_8$ that the 3-up-1-down spin configuration for each tetrahedron, same as in $ACr_2O_4$ [7,8], is realized and the tetrahedra are distorted to stabilize this spin configuration due to the strong spin–lattice coupling, even though there is a strong bond alternation indicated by the small $B_f$ of approximately 0.1. In addition, this is in contrast to the absence of the $M_s/6$ plateau in $LiGa_{0.125}In_{0.875}Cr_4O_8$, indicating that the spin-lattice coupling does not help to form the $M_s/6$ plateau.

Finally, we note the magnetization curve just below the half magnetization plateau. The magnetizations of $ZnCr_2O_4$ and $MgCr_2O_4$, which have large Weiss temperatures of −300 to −400 K, comparable to that of $LiInCr_4O_8$, linearly increase with increasing $H$ just below the half magnetization plateau [12,16]. In contrast, the *M–H* curve of $LiInCr_4O_8$ does not show such behavior, but shows a discontinuous jump to the half magnetization plateau. Note that a similar jump has been observed in the *M–H* curve of $CdCr_2O_4$ with a much weaker antiferromagnetic interaction ($\theta_W = -70$ K) [5]. Theoretically, the linearly increasing region preceding the plateau appears when the spin–lattice coupling is relatively weak compared to the strength of antiferromagnetic interaction [18]. Therefore, the spin–lattice coupling of this compound must be unusually strong compared to that in $ZnCr_2O_4$ and $MgCr_2O_4$. It is important to clarify the effect of bond alternation on the spin–lattice coupling in the breathing pyrochlore lattice.

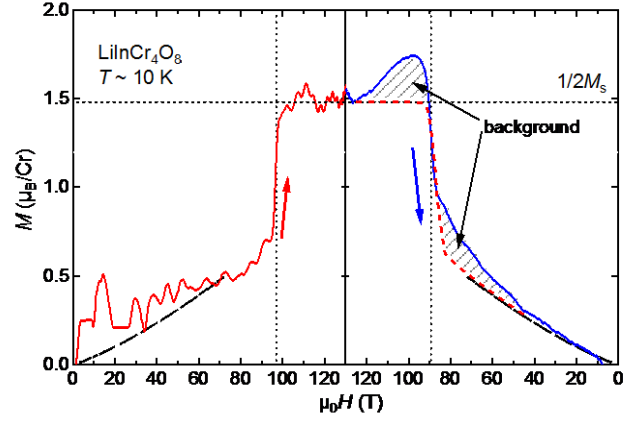

Fig. 3. Magnetization curves of a $LiInCr_4O_8$ powder sample obtained by a single-turn coil megagauss generator in magnetic fields of up to 130 T at approximately 10 K. The left- and right-hand-side panels present the *M–H* curves obtained in elevating and descending processes of a pulse magnetic field, respectively. The broken lines are the *M–H* curve measured by a nondestructive magnet at 1.4 K, as shown in Fig. 1(a). The horizontal dotted line indicates half the $M_s$. The shaded area is regarded as a background contribution increasing over time, which is caused by the time evolution of the inhomogeneity of the magnetic field in an expanding single-turn coil.

## IV. SUMMARY

We have reported the magnetization processes of spin-3/2 breathing pyrochlore antiferromagnets, measured up to 72 T by employing a nondestructive multilayered pulsed magnet and up to 130 T by the single-turn coil method. The magnetization of $LiInCr_4O_8$, which exhibits an antiferromagnetic order at 15 K, monotonically increases with increasing magnetic field up to 72 T. In contrast, the intrinsic magnetization of $LiGa_{0.125}In_{0.875}Cr_4O_8$, which shows a spin-gap-like magnetic susceptibility down to 2 K, is very small at a low magnetic field, while strongly increasing above 13 T, reflecting the closing of the spin gap between the tetramer singlet state and the excited state by applying a magnetic field. At approximately 90 T, the magnetization of $LiInCr_4O_8$ jumps to the half magnetization plateau, which continues to the highest measured field of $\mu_0 H = 130$ T. This result indicates the presence of considerably strong spin–lattice coupling in $LiInCr_4O_8$ and suggests that this compound is a unique system for realizing an intriguing magnetic property induced by strong spin–lattice coupling and bond alternation.

## ACKNOWLEDGMENTS

This work was mostly carried out at the International MegaGauss Science Laboratory under the Visiting Researcher Program of the Institute for Solid State Physics, Univ. of Tokyo and supported by JSPS KAKENHI Grant Number 16H03848. The authors are grateful to M. Mori and K. Takenaka for helpful discussions.

______________________________________________________

*yokamoto@nuap.nagoya-u.ac.jp